# A Robust Attack: Displacement Backdoor Attack

1st Yong Li
*School of Computer Science and Engineering, Changchun University of Technology*
*AI Research Institute, Changchun University of Technology*
*School of Computer Science and Technology, Jilin University*
Changchun 130012, China
liyong@ccut.edu.cn

2nd Han Gao
*School of Computer Science and Engineering*
*Changchun university of technology*
Changchun 130012, China
2202203116@stu.ccut.edu.cn

*Abstract*—As artificial intelligence becomes more prevalent in our lives, people are enjoying the convenience it brings, but they are also facing hidden threats, such as data poisoning and adversarial attacks. These threats can have disastrous consequences for the application of artificial intelligence, especially for some applications that take effect immediately, such as autonomous driving and medical fields. Among these threats, backdoor attacks have left a deep impression on people with their concealment and simple deployment, making them a threat that cannot be ignored, however, in the process of deploying the backdoor model, the backdoor attack often has some reasons that make it unsatisfactory in real-world applications, such as jitter and brightness changes. Based on this, we propose a highly robust backdoor attack that shifts the target sample and combines it with itself to form a backdoor sample, the Displacement Backdoor Attack(DBA). Experimental results show that the DBA attack can resist data augmentation that simulates real-world differences, such as rotation and cropping.

## I. INTRODUCTION

In recent years, deep neural networks (DNN) have been widely used in many important real-world such as recognition [1] Computer Vision [2] [3] [4] [5] [6], and machine translation [7]. Nonetheless, DNNs have been shown to be vulnerable to potential threats at multiple stages of their life cycle. such as data poisoning and adversarial attacks. These threats will have disastrous consequences for the application of artificial intelligence, especially some applications that take effect immediately, such as autonomous driving and medical fields. Among these threats, backdoor attacks have left a deep impression on people with their concealment and simple deployment, making them a threat that cannot be ignored. The common attack method of backdoor attacks is to force the target model to learn the trigger knowledge set by the attacker by changing part of the training data. The model that successfully learns this knowledge is called a backdoor model. When the trigger appears, these models containing backdoors will output a target category pre-specified by the attacker, which means that the attacker can make the model output the desired results according to his own needs, and these results often cause losses to people using AI. When the input does not contain a trigger, the backdoor model will behave like a normal model. Although backdoor attacks have become more and more advanced after years of development, the attack methods have become more diverse and are not limited to being implemented through poisoned data. However, most attacks still rely on triggers in the input samples, that is, relying on poisoned training samples to implement attacks. The most important thing is the setting of triggers. It is relatively easy to train the backdoor model by manipulating the training data. Especially in today's world where there is not much high-quality data, most model trainers will choose to train their models through third-party data sets. This process gives malicious third parties an opportunity to deploy backdoors. The core position of triggers in backdoor attacks is self-evident. The setting of triggers is crucial to the success of backdoor attacks. There are two factors that affect the effect of backdoor attacks: 1. The trigger setting of backdoor attacks should ensure high concealment to ensure that the backdoor data will not be processed by human selection before it is put into training. 2. The trigger setting should be simple enough, so that the target model can learn the backdoor knowledge and will not affect the decline of model accuracy as much as possible, in order to obtain an ideal backdoor model. Guided by the above two goals, we reproduced several classic backdoor attack algorithms. Through observation, we found that the early attacks set local triggers, such as Badnets [8], or global triggers, such as Blend [9], Refool [10], SIG [11], etc. The triggers set by these attacks are very obvious. On the one hand, it is because of the setting of the trigger, such as the distribution of the backdoor features, which is very different from the real distribution of the target sample. On the other hand, the trigger setting is fixed and will not change with the change of the input sample. People can easily distinguish the trigger of the backdoor sample. These

This work was supported by the Science and Technology Research Planning Project of Jilin Provincial Department of Education in China under Grant JJKH20230766KJ.

reasons make it very easy to pick out the backdoor samples carrying triggers by human eyes before model training. Even in the later more novel and covert attacks such as Wanet [12], SSBA [13], LIRA [14], etc., there are still some discordant noises in the backdoor samples. These noises destroy the characteristic continuity of the samples. That is, because of the appearance of these backdoor triggers, some pixels in the samples become more eye-catching, such as the irregular lines of the road signs in Wanet, and the snowflakes in SSBA and LIRA. Although these discordant factors are sufficiently hidden compared to the early backdoor attack algorithms, these backdoor samples can still be manually eliminated if the defenders have certain prior knowledge of backdoor attacks. In addition, no matter it is an early or newer algorithm, when deploying backdoors, the behaviors from target samples to backdoor samples are always similar. For example, in Badnets, backdoor samples all have white pixel blocks with the same position and feature value size, Blend, Refool, and SIG have the same interpolated background, and the attack algorithms in Wanet, SSBA, and LIRA all implement the same backdoor behavior on different samples. The only difference is whether the backdoor triggers generated after the backdoor behavior is implemented are consistent. From a visual point of view, it means whether the difference before and after the change of different backdoor samples is equal. If they are equal, such as Badnets, Blend, etc., and if they are not equal, such as Wanet, SSBA, etc. Through this feature, we can guess that backdoor attacks can be easily deployed when the backdoor attacks meet two assumptions: 1. Choosing an appropriate and consistent backdoor behavior can turn benign samples into backdoor samples, that is, the steps of setting the trigger are consistent. 2. The set trigger can be learned by the target model. In addition, the model is very rigid in learning triggers, that is, in the deployment phase, when the triggers contained in the input samples deviate from the attacker's pre-set triggers, such as the location and feature values of the triggers, it is difficult for the backdoor model to output the target class specified by the attacker, especially for those backdoor attack algorithms that are not sensitive to input. Based on the two assumptions mentioned above, in this section we propose a backdoor attack algorithm with high concealment. Compared with the previous attacks, the setting of the backdoor trigger in our attack algorithm is input-aware and simpler. This attack algorithm uses the displacement of the target sample as a trigger, and merges the displaced sample with the original sample to form a backdoor sample, We call this attack method a displacement backdoor attack(DBA). Visually, the backdoor sample is similar to the afterimage left by pressing the shutter on an object when the camera moves. This afterimage is the trigger of the algorithm. The attack example of the algorithm is shown in FIG.

Nonetheless, DNNs have been shown to be vulnerable to potential threats at multiple stages of their life cycle. In reality, users often use data sets provided by third parties to train their models for some reasons. This also gives malicious parties the opportunity to deploy backdoors in the model by training it on data which provided by adversary. Intuitively, backdoor attacks aim to trick the model into learning strong correlations between trigger patterns and target labels by poisoning a small portion of the training data. Backdoor attacks can be notoriously dangerous for several reasons. First, backdoor data can infiltrate models in many situations, including training models on data collected from untrustworthy sources or downloading pre-trained models from untrusted parties. Existing defense methods can be roughly divided into two categories based on samples: one is a method that requires additional clean samples, and uses clean samples to fine-tune [15], prune [16] [17] [18] or other operations [19] [20]to eliminate or reduce the impact of backdoor attacks. The other is a method that does not require additional samples for defense, such as Anti-Backdoor-Learning(ABL) [21], data augmentation [22] [23] and distillation [24] [19].

In view of the problems of obvious triggers and abnormal pixels in traditional backdoor attack algorithms, this chapter proposes an attack algorithm that uses the input sample itself as a trigger, which further improves the concealment of the trigger and reduces the possibility of human elimination. On this basis, we also verified that the DBA attack can cope with ASR changes affected by changes in the real world, and found that our algorithm has good robustness, which means that even in reality, even if the model that has been attacked by DBA receives input that is somewhat different from the training data, it can still output the target class specified by the attacker in a relatively stable manner. Through experiments, it can be observed that the attack and defense efficiency of the backdoor attack algorithm proposed in this chapter reaches the optimal or suboptimal results in most cases, and its robustness detection results also maintain a high level, which can cope with some changes brought about by the physical world.

## II. RELATED WORK

Trigger: In the beginning, adversary training a backdoored modules by adding patches [8] or multiple and scatter pixel [25] [13] to a part of the samples in the training data set, make the model learns the knowledge of the backdoor trigger and changes its decision boundary. However, this form of trigger is not covert, so more stealthier triggers were proposed in later backdoor attacks, such as the BLEND [9], a method of injecting triggers through picture interpolation, then the Sinusoidal signal attack (SIG) [11], Reflection attack (Refool) [10] and Convex Polytope Attack [26] is proposed. As well as some attack methods that use items in reality as triggers [27]. Some recent attack methods have generated triggers that are difficult to detect by the human eye, such as Wanet [12] and LIRA [14]. These more covert backdoor attacks have caused great trouble for defense.

Poisoning methods: Poisoning methods can be roughly divided into two types depending on whether to change the ground truth of the poisoned sample. A method of changing the poisoning sample to the target class specified by the attacker to complete the backdoor attack, such as BadNet [8], we call this method dirty label attack. The other is to keep the label of the poisoned sample consistent with the original label, such as SIG [11] and label-consistent (LC) [28]. We call this method clean label attack. This method avoids the possibility of being detected to a certain extent.

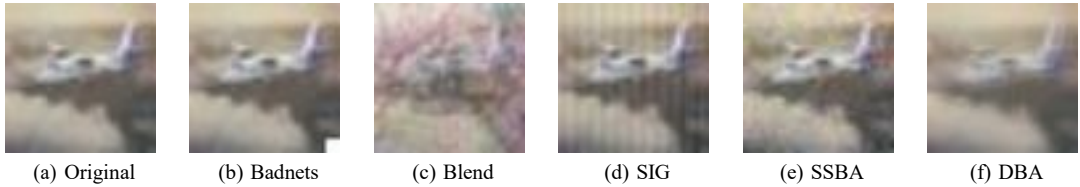

(a) Original    (b) Badnets    (c) Blend    (d) SIG    (e) SSBA    (f) DBA

Fig. 1: Figure a is the original image, and Figures b,c,d,e,f show examples of poisoned samples of various backdoor attacks.

Defense: Here we only classify defense methods according to whether they require additional samples. First, defense methods that require additional samples such as fine-tuning [29] [18], retraining, pruning [16] [17] [16], clustering [30], or defense methods that use distillation that require additional samples [20]. The other type does not require additional samples and is defended through the characteristics of backdoor attacks, such as ABL [21] that isolates toxic samples based on the loss function, and combat backdoor through data augmentation like Strong data augmentation [23], Deepsweep [22], and novel way such as model connectivity repair (MCR) [31],Neural Attention Distillation (NAD) [19]. There are also some defense methods based on recurrence triggers among them [32] [33] [34] [35] [36]

## III. METHOD

In this chapter, we will introduce the overall method flow and the meaning of symbols and the entire attack process.

The backdoor attack algorithm we proposed is very simple. You only need to complete the following steps to implement the backdoor attack: 1. First, determine the category to be attacked. 2. Select a certain proportion (here 90%) of the samples in the analogy as the target for deploying the trigger. 3. Displace the selected samples several times, and use the displaced samples as spare samples, and finally interpolate with the sample. The formula is as follows:

$$(1 - N\alpha)x + \alpha * x_1 + \alpha * x_2 ... \alpha * x_n = x_t \quad (1)$$

where x is the selected target sample, $x_n$ is the sample carrying the trigger, y is the category corresponding to x, that is, the backdoor sample, D () is the backdoor generator of the displacement backdoor attack, and $\alpha$ is a parameter that controls the obviousness of the trigger.

## IV. EXPERIMENT

In this chapter, we will present the settings of our entire experiment, such as learning rate, data poisoning ratio of backdoor attacks, and the results of the experiment

### A. Dataset

(1) Cifar-10 [?] witch has 10 categories, each containing a total of 5000 samples, totaling 50000 samples. We conducted two poisoning methods. (I) Using a non label clean approach, 250 samples were randomly selected from each category, with a total of 2500 samples accounting for 5% of the whole dataset, and placed in the target category called bird in this paper. (II) Using a label clean method, randomly select 2500 samples from the target label, accounting for 5% of the total dataset for poisoning.

(2) Mnist [?].The MNIST dataset is from the National Institute of Standards and Technology (NIST) in the United States. The training set consists of handwritten numbers from 250 different individuals, of which 50% are high school students and 50% are staff from the Census Bureau. The test set also has the same proportion of handwritten digit data, but ensures that the author set of the test set and the training set do not intersect Two methods of poisoning were used. (I) Using a non label clean approach, 10% samples were randomly selected from each category and add them into the target category, called number 3 in this paper. (II) Using a label clean method, randomly select half of samples from the target label.

### B. Attack Setting

Attacker: As the implementer of the backdoor attack, you can freely change the feature values and distribution of the training data, that is, the setting of the backdoor trigger. Purpose: 1. Ensure that the difference in feature distribution between the backdoor sample and its corresponding original sample is not so large that it can be excluded by the human eye. 2. Train a backdoor model $M_b$ that outputs the category $Y_t$ specified by the attacker when the trigger $\alpha$ pre-set by the attacker appears, and outputs the correct category when the trigger does not appear, as shown in the formula. 3. Ensure that the output accuracy of the model for normal data is not much different from that of the benign model and the attack success rate is as high as possible.In this chapter, the experiment uses Resnet-50 as the target model, and implements four representative backdoor attack methods, BadNets, SIG, Blend, and SSBA, as comparative attack algorithms to evaluate the effect of DBA on the attack. The above attack algorithms are applied to the image classification datasets CIFAR-10 and MNIST respectively. The attack effect is shown in the figure. The attacker uses five attack methods, BadNets, Blend, SIG, SSBA, and DBA, to train the backdoor model on two datasets. The attack settings are as follows: 1. BadNets: On the MNIST and CIFAR10 datasets, a 2x2 white block is covered in the lower right corner of the poisoned data, as shown in the figure. The attack method is a label consistency attack, and the total amount of poisoned data is 1% of the total amount of training data. 2. Blend: On the MNIST and CIFAR10 datasets, the

$(1-N\alpha)* \text{[image]} + \alpha_1 * \text{[image]} + \alpha_2 * \text{[image]} + \ldots\ldots = \text{[image]}$

Fig. 2: Stream Line: This figure demonstrates the entire defense method process

TABLE I: The data in this table are all taken from the last round results of various defense training, among which the best result is displayed in bold in bold

| Backdoor Attack | Baseline | | ABL | | NAD | | SPT | | FT | | FT-sam | | FST | |
|---|---|---|---|---|---|---|---|---|---|---|---|---|---|---|
| | ACC↑ | ASR↓ | ACC↑ | ASR↓ | ACC↑ | ASR↓ | ACC↑ | ASR↓ | ACC↑ | ASR↓ | ACC↑ | ASR↓ | ACC↑ | ASR↓ |
| BadNet | 98 | 100 | 98.4 | 0.0 | 89.7 | 0.3 | 97.5 | 0.1 | 99 | 99.7 | 96.1 | 87.2 | 98.2 | 0.0 |
| Blend | 98.8 | 100 | 79.1 | 24.5 | 83.2 | 46 | 84.2 | 45.8 | 75.5 | 85.6 | 89.8 | 99 | 99 | 85.8 |
| SIG | 98.2 | 100 | 58.1 | 100 | 83.4 | 11.8 | 83.3 | 6.1 | 89.8 | 99.7 | 90.3 | 100 | 98.9 | 42 |
| SSBA | 99.4 | 100 | 13.5 | 1.4 | 89.4 | 9.8 | 97.3 | 10 | 98.5 | 10 | 99 | 10 | 99 | 10.1 |
| DBA-LC | 89.3 | 100 | 16.3 | 99.2 | 88.9 | 0.2 | 97.4 | 0.3 | 99.1 | 16.8 | 99 | 94.3 | 98.7 | 0.2 |
| DBA-DL | 98.4 | 100 | 98.2 | 5.7 | 89.4 | 0.3 | 90.7 | 0.3 | 98.9 | 100 | 98.9 | 100 | 98.4 | 0.1 |

TABLE II: !!!

| Backdoor Attack | Baseline | | ABL | | NAD | | SPT | | FT | | FT-sam | | FST | |
|---|---|---|---|---|---|---|---|---|---|---|---|---|---|---|
| | ACC↑ | ASR↓ | ACC↑ | ASR↓ | ACC↑ | ASR↓ | ACC↑ | ASR↓ | ACC↑ | ASR↓ | ACC↑ | ASR↓ | ACC↑ | ASR↓ |
| BadNet | 79.4 | 96.3 | 78.3 | 1.1 | 65.9 | 10.4 | 79.8 | 11.1 | 75.8 | 13.4 | 80.1 | 13.7 | 74.7 | 8.9 |
| Blend | 79.9 | 92.6 | 75.8 | 81.8 | 70.5 | 12.7 | 73.6 | 37.7 | 73.6 | 20.6 | 79.5 | 5.8 | 71.3 | 51.8 |
| SIG | 79.5 | 97.6 | 75.3 | 5.0 | 55.6 | 1.2 | 77.1 | 12.4 | 74.3 | 2.4 | 80.5 | 8.0 | 75 | 4.5 |
| SSBA | 79.1 | 99.8 | 78.8 | 18.5 | 68.8 | 9.2 | 76 | 7.5 | 67.1 | 9.2 | 83.9 | 34.2 | 79.8 | 3.4 |
| DBA-LC | 74.3 | 92.5 | 65.8 | 73.4 | 67.5 | 9.2 | 74.8 | 19.9 | 75.3 | 10.1 | 79.4 | 12.1 | 77.7 | 11.3 |
| DBA-DL | 80.3 | 98.6 | 76.3 | 0.8 | 68.7 | 13.5 | 75.7 | 20.4 | 70.1 | 12.8 | 82.3 | 25.1 | 77.4 | 11.6 |

trigger size is set to be consistent with the input data, the trigger style is set as shown in the figure, the interpolation (here represented by a symbol) is 0.3, the attack method is a label consistency attack, and the total amount of poisoned data is 1% of the total amount of training data. 3.SIG: On the MNIST and CIFAR10 datasets, the trigger size is set to be consistent with the input data, the interpolation is 0.03, the attack method is label consistency attack, and the total amount of poisoned data is 1% of the total training data. 4.SSBA: adopt label consistency attack, 5.DBA: adopt label consistency attack, the attack ratio is 70% of the target sample category, which is reflected in MNIST and CIFAR-10, accounting for about 9% and 7% of the total data respectively. In the case of non-label consistency attack, 0.1% of the total data is poisoned.

### C. Defence Setting

Defender: The defender has all the rights in the process of training the model, including but not limited to the selection of the model, the change of the model structure, the adjustment of the learning rate and the loss function, and also retains the right to change the data set, including data enhancement methods such as rotation and cropping. The goal of the defender is to reduce the success rate of backdoor attacks on the model, maintain the accuracy of the model for normal input, and ensure that the model can be used normally.

The DBA method is compared with six defense methods in the field of backdoor defense: 1. Fine-tuning (FT) [68]. 2. Neural Attention Distillation (NAD) [56]. 3. Anti-backdoor Learning (ABL). 4. SPT, 5. Sharpness-aware fine-tuning (FT-SAM), 6. Feature Shift Tuning (FST). In the Mnist data, FT, FT-SAM, SPT, NAD, and FST set the number of benign data sets to 10%, 10%, 20%, 10%, and 0.1% of the original data respectively. The purpose is to verify the performance of the DBA method in backdoor defense, and to analyze the differences in the attack effects of the DBA method by setting up multiple groups of comparative experiments.

### D. Evaluation Metrics

In terms of measurement methods, we adopted Attack Success Ratio (ASR) and Accuracy (ACC). The first method measures how many of the test sets with triggers are judged as the target class. To avoid errors, we remove all samples of the corresponding target class in the test set when testing ASR. The second method gives the performance of the model on the normal data set, which is a traditional measurement method.

### E. Result

The robustness test result graph, Figure 1, shows the attack success rate of each round of training of the backdoor model. It can be observed from the robustness test result line graph that in MNIST, the robustness ASR of the DBA algorithm has maintained a high level since the beginning of training, and gradually stabilized with the increase of ASR. Compared with other attacks, the attack success rates of $DBA_DL$ and $DBA_LC$ remained stable and close to 100% after experiencing random data enhancement combinations, while the attack success rates of other comparison algorithms Badnets,

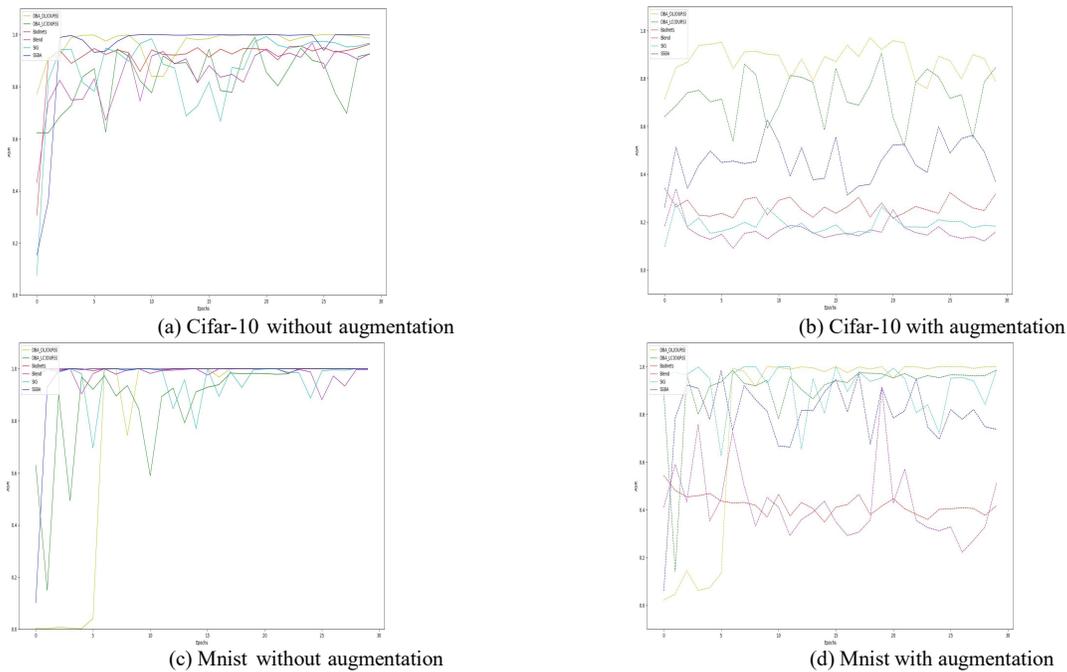

Fig. 3: a, b, c, and d show the attack performance of backdoor attacks based on CIFAR-10 and MNIST datasets without data augmentation and with data augmentation, respectively.

Blend, and SIG were less than 60% for a long time after random data enhancement, while SSBA remained relatively stable in the face of data enhancement, and ASR remained around 80%. In CIFAR-10, whether it is label consistency or non-label consistency, the ASR stability of DBA facing data enhancement is far inferior to MNIST, but compared with other attack algorithms mentioned in this article, it still maintains a relatively high level. The ASR is stable at around 75%, while the robustness ASR of Badnets, Blend, and SIG is stable at around 20%. Among them, SSBA performs best, with an attack success rate of around 43%.

## V. CONCLUSION

In view of the obvious triggers and abnormal pixels in traditional backdoor attack algorithms, this chapter proposes an attack algorithm that uses the input sample itself as a trigger, which further improves the concealment of the trigger and reduces the possibility of human elimination. This method first determines the attack category and randomly selects samples based on 90% of the target class or selects 0.1% of the samples under the condition of non-label consistency. The trigger is obtained by simply shifting the original sample and merging it with the original sample to obtain a poisonous data set. The backdoor model trained based on this data set can achieve more than 90% or nearly 100% ASR. On this basis, we also verified that DBA attacks can cope with ASR changes affected by changes in the real world and found that our algorithm has good robustness, which means that even in reality, even if the model that has been attacked by DBA receives a certain difference in input from the training data, it can still output the target class specified by the attacker relatively stably. Through experiments, it can be observed that the attack and defense efficiency of the backdoor attack algorithm proposed in this chapter reaches the optimal or suboptimal results in most cases, and its robustness detection results also maintain a high level, which can cope with some changes brought by the physical world. However, the triggers created by the DBA algorithm have defects to a certain extent. For example, the brightness of the original image is different, and some images appear blurry. This is also one of the problems we need to overcome in the future.
## REFERENCES

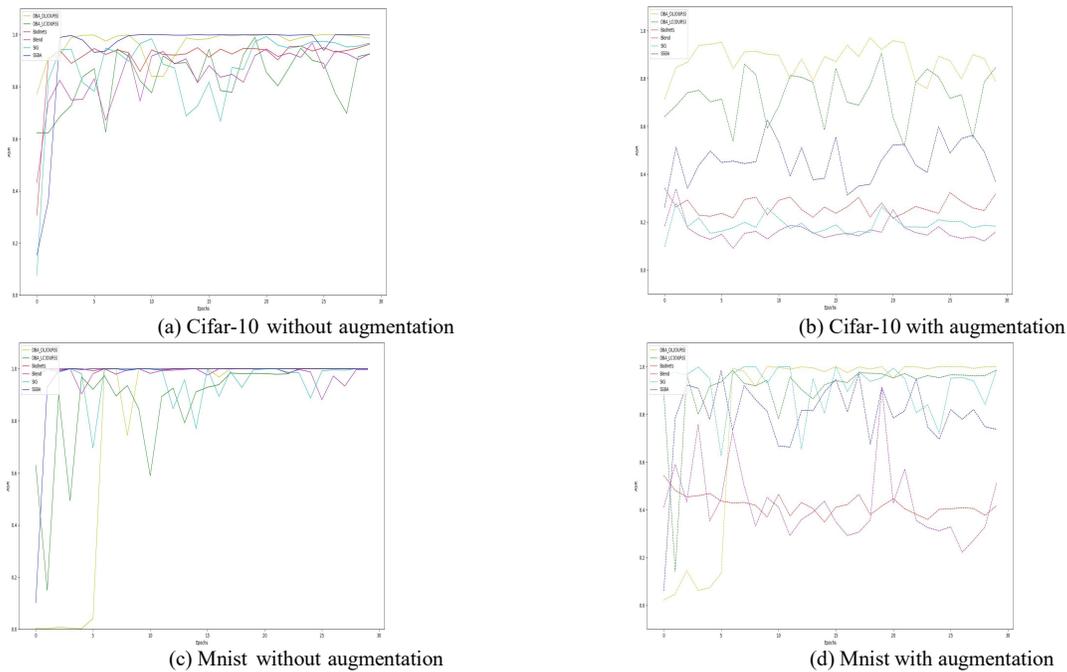

Fig. 3: a, b, c, and d show the attack performance of backdoor attacks based on CIFAR-10 and MNIST datasets without data augmentation and with data augmentation, respectively.

Blend, and SIG were less than 60% for a long time after random data enhancement, while SSBA remained relatively stable in the face of data enhancement, and ASR remained around 80%. In CIFAR-10, whether it is label consistency or non-label consistency, the ASR stability of DBA facing data enhancement is far inferior to MNIST, but compared with other attack algorithms mentioned in this article, it still maintains a relatively high level. The ASR is stable at around 75%, while the robustness ASR of Badnets, Blend, and SIG is stable at around 20%. Among them, SSBA performs best, with an attack success rate of around 43%.

## V. CONCLUSION

In view of the obvious triggers and abnormal pixels in traditional backdoor attack algorithms, this chapter proposes an attack algorithm that uses the input sample itself as a trigger, which further improves the concealment of the trigger and reduces the possibility of human elimination. This method first determines the attack category and randomly selects samples based on 90% of the target class or selects 0.1% of the samples under the condition of non-label consistency. The trigger is obtained by simply shifting the original sample and merging it with the original sample to obtain a poisonous data set. The backdoor model trained based on this data set can achieve more than 90% or nearly 100% ASR. On this basis, we also verified that DBA attacks can cope with ASR changes affected by changes in the real world and found that our algorithm has good robustness, which means that even in reality, even if the model that has been attacked by DBA receives a certain difference in input from the training data, it can still output the target class specified by the attacker relatively stably. Through experiments, it can be observed that the attack and defense efficiency of the backdoor attack algorithm proposed in this chapter reaches the optimal or suboptimal results in most cases, and its robustness detection results also maintain a high level, which can cope with some changes brought by the physical world. However, the triggers created by the DBA algorithm have defects to a certain extent. For example, the brightness of the original image is different, and some images appear blurry. This is also one of the problems we need to overcome in the future.